\documentclass[final,3p,times,twocolumn]{elsarticle}
\usepackage{graphics}
\usepackage{epsfig}

\biboptions{}

\journal{Physics Letters B}

\begin{document}

\begin{frontmatter}

%% Title, authors and addresses

%% use the tnoteref command within \title for footnotes;
%% use the tnotetext command for the associated footnote;
%% use the fnref command within \author or \address for footnotes;
%% use the fntext command for the associated footnote;
%% use the corref command within \author for corresponding author footnotes;
%% use the cortext command for the associated footnote;
%% use the ead command for the email address,
%% and the form \ead[url] for the home page:
%%
\title{CMS ridge effect at LHC as a manifestation of bremstralung of gluons due to the quark-anti-quark string formation}
%% \tnotetext[label1]{}
\author{B.A. Arbuzov, E.E. Boos and V.I. Savrin}
%% \ead{arbuzov@theory.sinp.msu.ru}
%% \ead[url]{home page}
%% \fntext[label2]{}
%% \cortext[cor1]{}
\address{Skobeltsyn Institute of Nuclear Physics of 
MSU,  119992 Moscow, RF}
%% \fntext[label3]{}

\title{CMS ridge effect at LHC as a manifestation of bremstralung of gluons due to the quark-anti-quark string formation}

%% use optional labels to link authors explicitly to addresses:
%% \author[label1,label2]{<author name>}
%% \address[label1]{<address>}
%% \address[label2]{<address>}

%\author{}

%\address{}

\begin{abstract}
The recently reported effect of long-range near-side angular correlations at LHC occurs for large multiplicities of  particles with $1\,GeV\,<p_T\,<\,3\,GeV$. To understand the effect several possibilities have been discussed. In the letter we propose a simple qualitative mechanism which  corresponds to 
gluon bremstralung of quarks moving with acceleration appropriate to the quark--anti-quark string. The smallness of  
azimuthal angle difference $\Delta \phi$ along with large $\Delta \eta$ at large multiplicities in this interval of $p_T$ are natural in the mechanism. The mechanism predicts also bremstralung photons with mean values of $p_T \approx 2.9$ and $0.72\,GeV$. 
\end{abstract}

\begin{keyword}
CMS ridge effect \sep quark-anti-quark string
\sep gluon bremstralung
%% MSC codes here, in the form: \MSC code \sep code
%% or \MSC[2008] code \sep code (2000 is the default)

\end{keyword}

\end{frontmatter}

%%
%% Start line numbering here if you want
%%
% \linenumbers

In paper~\cite{CMS} the effect in proton-proton collisions at LHC is reported for existence of a ridge in the plot of data 
for two-particle correlations versus pseudo-rapidity difference $\Delta \eta$ and azimuthal angle $\Delta \phi$ plane. This ridge means essential excess of events with $\Delta \phi$ close to zero and large $\Delta \eta$. It is important to emphasize, that the effect is observed under condition that the accompanying charged particles have multiplicity $>100$ and are each situated in restricted region of transverse momentum $1\,GeV\,<p_T\,<\,3\,GeV$.

The result already causes discussion devoted to possible interpretation of the data~\cite{Shur} -- \cite{Troshin}. In note~\cite{Shur} possibilities of interpretation of the effect in terms of quark-anti-quark strings are discussed. The first point is that a string is formed being stretched close to the direction of $p\,p$ collision and  thus might decay with $\Delta \phi$ $\approx 0$ and $\approx \pi$. However it is emphasized 
in~\cite{Shur}, that there is no reason, why 
the effect is observed only for very high multiplicity $N > 100$, while multiple production of hadrons via string decomposition leads to randomization and thus simulations show no ridge.
 
In the present letter we would try to explain why even for high multiplicity the correlation persists when we take into account radiation of gluons in the process of QCD quark--anti-quark string formation in proton-proton collision at very high energies. We mean formation of a string between either quark from the first proton and anti-quark from the second one or vice versa. The direction of the string is evidently close to the direction of momenta of the colliding protons. The string has overall motion with some momentum.

Each quark (anti-quark)  moves considerable time in a very strong color field which we 
describe in terms of the string tension. In this case an  acceleration has almost the same direction as the velocity and thus we may use the well-known classical expression for dipole electromagnetic radiation of electric charge $e$ in electric field parallel to velocity of the motion~\cite{LaLi}
\begin{equation} 
\frac{dE}{dt}\,=\,\frac{2\,\alpha}{3}\,\Biggl(\frac{\mathbb A^2}{m}\Biggr)^2\,;\label{LL}       
\end{equation}    
where $m$ is a light quark mass, $\mathbb A^2$ is the string tension and $\alpha$ is the fine structure constant. We take initial expression~(\ref{LL}) in an accompanying reference frame. In view to make estimates we take the following values for these fundamental quantities
\begin{equation}
m_u = 2.5\,MeV ;\;  m_d = 5\,MeV ;\; \mathbb A = 420\,MeV ;\label{am}
\end{equation}
where light quark masses are chosen to be in the middle of interval of their possible values: $1.7\,MeV\,<\,m_u\,<3.3\,MeV\,;\;4.1\,MeV\,<\,m_d\,<5.8\,MeV$~\cite{PDG}.         
So quarks are moving with acceleration and thus radiate gluons. Let us obtain simple quasi-classical estimate of the mean energy of radiated gluons. In view of this we rewrite expression~(\ref{LL}) in the form 
\begin{equation}
\frac{\Delta E}{\Delta t}\,=\,\frac{\alpha_s}{9}\,\Biggl(\frac{\mathbb A^2}{m}\Biggr)^2\,;\label{delta}         
\end{equation}
where we change in~(\ref{LL})  
$\alpha \to \alpha_s$ and introduce the evident color factor. 
Now to obtain the quasi-classical estimates we use the well-known uncertainty relation
\begin{equation}
\Delta E\,\Delta t\,=\,1\,.\label{ET}         
\end{equation}
Finally we have for the mean energy of a gluon 
\begin{equation}
\Delta E\,=\,\sqrt{\frac{\alpha_s}{9}}\,\frac{\mathbb A^2}{m}\,.\label{DE}         
\end{equation}
Then we use the standard one loop expression for $\alpha_s$ at scale $\Delta E$
\begin{equation}
\alpha_s(\Delta E)\,=\,\frac{12 \pi}{(33-2 N_f)\,\ln\,\Bigl(\frac{\Delta E^2}{\Lambda_{QCD}^2}\Bigr)}\,;\label{alphas}         
\end{equation}
We have for one loop expression~(\ref{alphas}) with $N_f = 4$, $\Lambda_{QCD}\,\approx\,190\,MeV$\footnote{We normalize $\alpha_s$ at the point of $\tau$-lepton mass due to better precision of data here.}. Then with this result the solution of 
relations~(\ref{DE}, \ref{alphas}) under conditions~(\ref{am}) gives us the following estimates for radiation off quarks $u$ and $d$ 
\begin{equation}
\Delta E_u\,\approx\,11.2\,GeV\,;\quad \Delta E_d\,\approx\,5.6\,GeV\,.\label{DER}         
\end{equation}
The result~(\ref{DER}) gives an estimate for mean energies of the bremstralung gluons. 

First of all let us consider  an explanation of large differences in pseudo-rapidity $\Delta\eta$ along with small differences in azimuthal angle $\Delta\phi$. Here we are to take into account both quarks constituting the string. Namely let the string be produced with some overall momentum $k$ while its direction remains being (almost) parallel to the line of $p\,p$ collision. Such situation is presented in Fig.1.
\\ 
\\
\\
\begin{picture}(180,30)
{\thicklines
\put(20,5){\vector(1,0){80}}
\put(180,5){\vector(-1,0){80}}
\put(50,8){\line(1,0){100}}
\put(50,8){\circle*{3}}
\put(150,8){\circle*{3}}
\put(100,8){\vector(-1,3){8}}
\put(100,8){\vector(-1,0){56}}
\put(100,8){\vector(1,0){56}}
%\put(30,8){\vector(-1,3){4}}
%\put(90,8){\vector(-1,3){4}}
\put(100,8){\vector(-3,1){68}}
\put(100,8){\vector(2,1){52}}
\put(100,30){k}
\put(50,-5){p}
\put(60,12){$\psi_1$}
\put(130,12){$\psi_2$}
\put(150,-5){p}}
\put(60,27){$q_1$}
\put(120,30){$q_2$}
\end{picture}
\begin{flushleft}
Fig.1. The string moving with momentum $k$ from the point of collision of two protons, $\psi_1,\,\psi_2$ are angles in Eq.(\ref{LLthetapsi}) and $q_1$, $q_2$ are momenta of the quarks.\\
\end{flushleft}
Then 
velocities of quarks are not parallel to the direction of acceleration, but constitute some angles $\psi_1, \psi_2$ with this direction. When a velocity and an acceleration are not parallel {\bf va}$\,=\,v\,a\cos\psi$ and there are two accelerated quarks we have 
the following angular distribution~\cite{LaLi}
\begin{eqnarray}
& &\frac{dE}{dt'}\,=\,\frac{\alpha_s}{24\,\pi}\,
\Biggl(\frac{\mathbb A^2}{m}\Biggr)^2\,\times\nonumber\\
& &\Biggl(\Phi(\psi_1,\theta,\phi,v_1)+\Phi(\psi_2,\theta,\phi,v_2)\Biggr)\,d\,\Omega;\label{LLthetapsi} \\
& &\Phi(\psi,\theta,\phi,v)=\frac{X\,+\,v^2 Y}{Z^5}\nonumber\\
& &X\,=\,\sin^2\theta-2 v \sin\psi \sin\theta \cos\phi\nonumber\\
& &Y\,=\,\cos^2\theta \sin^2\psi + \sin^2\theta \sin^2\psi \cos^2\phi\nonumber\\
& &Z\,=\,1-v(\cos\psi \cos\theta+\sin\psi \sin\theta \cos\phi);\nonumber
\end{eqnarray}
where $t'$ is a time with account of a retardation~\cite{LaLi}, $\psi_1,\,\psi_2$ are respectively angles for the first and the second quark. Small $\Delta\phi$ along with the wide spread in pseudo-rapidity in the effect~\cite{CMS} is connected with the same sign of $\sin\psi$, because quarks are directed to one side from the line of collision (see Fig.1). Then the distribution in polar angle $\theta$ has two maxima divided by some significant interval $\Delta \theta$, while distribution in azimuthal angle $\phi$ is again close to zero. After integration 
 of~(\ref{LLthetapsi}) by $\phi$ and $\theta$ correspondingly we have these distributions. The situation is illustrated in Fig.2 and Fig.3, in which we present normalized distribution in rapidity $\eta$ and normalized   
angular distribution in $\phi$ in $\phi$ for $\psi_1=0.1,\,\psi_2=\pi-0.1$,  $v_1=v_2=0.999$.
\\
\\
\\
\\
\begin{figure}[ht]
\begin{picture}(150,200)
\put(20,0){\epsfig{file=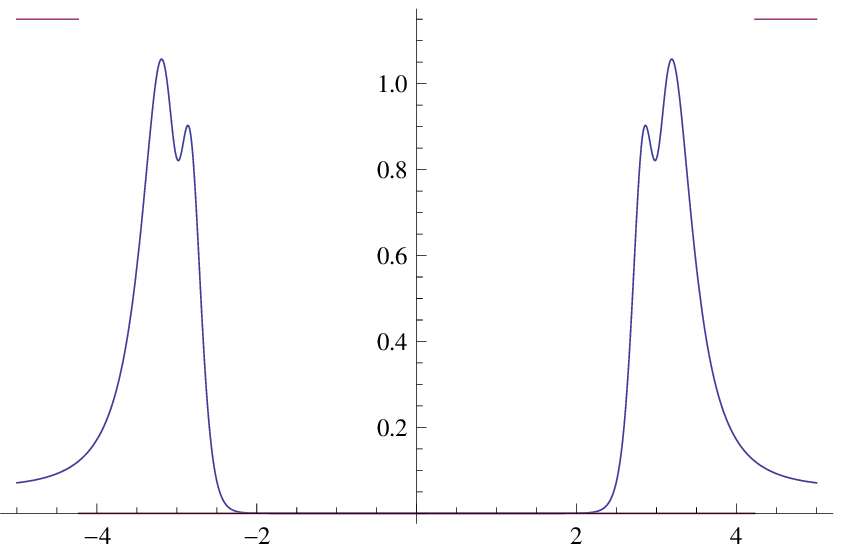,width=150pt, height=250pt}}
                         % this instruction read file (s_chain.eps)
                         % end define size of picture in pt (width/height)
%\put(90,10){Fig. 2}
\put(170,0){\Large {$\eta$}}
\put(60,220){$\Phi_N(\eta)$}
\end{picture}
\end{figure}
\begin{flushleft}
Fig.2. Behaviour of  $\Phi_N(\eta)$, $v=0.999$, $\psi_1=0.1$, $\psi_2=\pi-0.1$, for $-5 < \eta < 5$.\\
\end{flushleft}
\begin{eqnarray}
& &\frac{dE(\eta)}{dt'}\,=\,\frac{\alpha_s}{24\,\pi}\,\Biggl(\frac{\mathbb A^2}{m}\Biggr)^2\,\Phi(\eta)\,\frac{d\,\eta}{\cosh^2\eta};\nonumber\\
& &\Phi(\eta)\,=\,\int_{-\pi}^\pi\,\Phi_{12}(\psi_i, v_i,\theta,\phi)\,_{\cos \theta\,=\,f(\eta)}\,d\phi\,;\label{LLthpsi}\\
& &f(\eta)\,=\,\frac{\sinh \eta}{\cosh \eta}\,;\nonumber\\
& &\frac{dE(\phi)}{dt'}\,=\,\frac{\alpha_s}{24\,\pi}\,\Biggl(\frac{\mathbb A^2}{m}\Biggr)^2\,\Phi(\phi)\,d\phi\,;\nonumber\\
& &\Phi(\phi)\,=\,\,\int_0^\pi\,\Phi_{12}(\psi_i, v_i,\theta,\phi)\,\sin \theta\,d\,\theta\,;\label{LLphipsi}\\
& &\Phi_{12}(\psi_i, v_i,\theta,\phi)= \Phi(\psi_1,\theta,\phi,v_1)+\nonumber\\
& &\Phi(\psi_2,\theta,\phi,v_2).\nonumber
\end{eqnarray}
\begin{figure}[ht]
\begin{picture}(150,180)
\put(20,0){\epsfig{file=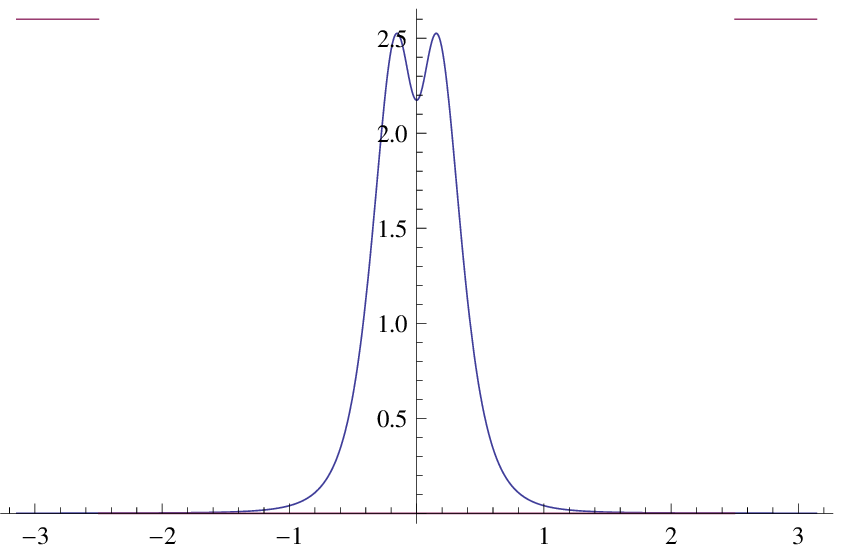,width=150pt,
height=200pt}}
                         % this instruction read file (s_chain.eps)
                         % end define size of picture in pt (width/height)
%\put(90,10){Fig. 3}

\put(175,5){\Large {$\phi$}}
\put(55,180){$\Phi_N(\phi)$}
\end{picture}
%\label{}
\end{figure}
\begin{flushleft}
Fig.3. Behaviour of  $\Phi_N(\phi)$, $v=0.999$,  $\psi_1=0.1$, $\psi_2=\pi-0.1$, for $-\pi < \phi < \pi$.\\
\end{flushleft}
From Fig.2, Fig.3 we see that $\Delta \eta$ may be quite significant while $\Delta \phi$ is small. One should note that the peaks in Fig.2 and Fig.3 become narrower
with increasing of speed and with increasing of $\psi$. 

Now let us consider properties of gluon radiation of a single quark. For this purpose we approximately assume the same direction of the velocity and of the acceleration in the actual reference frame of LHC. Angular distribution is described by the following expression~\cite{LaLi}
\begin{eqnarray}  
& &\frac{dE}{dt'}\,=\,\frac{\alpha_s}{24\,\pi}\,\Biggl(\frac{\mathbb A^2}{m}\Biggr)^2\,\frac{\sin^2\theta'}{(1-v \cos \theta')^5}\,d\,\Omega\,=\nonumber\\
& & \frac{\alpha_s}{24\,\pi}\,\Biggl(\frac{\mathbb A^2}{m}\Biggr)^2 \Phi_0(\theta')\,d\,\Omega;\label{LLtheta}\\
& &\Phi_0(\theta')\,=\,\Phi(0,\theta',\phi,v)\,;\nonumber   
\end{eqnarray}
where $v$ is a velocity of a quark, $\theta'$ is a polar angle and $d \Omega\,=\,\sin \theta' d \theta'\,d \phi$. Using angular distribution of the radiation~(\ref{LLtheta}) we estimate the mean $p_T$ of the radiated gluon
\begin{eqnarray}
& &<p_T^g> = \frac{\Delta E}{\sqrt{1-v^2}} \frac{I_1}{I_2}\,;\quad 
I_2 = \int \Phi_0(\theta')\, d\Omega\,;\nonumber\\
& &I_1\,=\,\int \Phi_0(\theta')\,A(v,\theta')\sin \theta'\, d\Omega\,;\label{int}\\
& &A(v,\theta')=1\,+\,\frac{\cos \theta'\, (1 - v^2) - v \sin^2 \theta'}{1 - v^2 \cos^2 \theta'}\,;\nonumber
\end{eqnarray}
where $\Phi_0(\theta')$ is defined in (\ref{LLtheta}). 
Calculating integrals in~(\ref{int}) with the aid of the following relation valid for $v\,\to\,1$ and $\rho > \frac{\mu}{2}$
$$
\int_0^\pi\frac{\sin^{\mu-1}\theta d \theta}{(1-v\,\cos\theta)^\rho} = \frac{2^{\mu-\rho} \Gamma\Bigl(\frac{\mu}{2}\Bigr)\Gamma\Bigl(\frac{1}{2}\Bigr) \Gamma(2 \rho-\mu)}{(1-v^2)^{\rho-\mu/2} \Gamma(\nu) \Gamma\Bigl(\frac{1-\mu}{2}+\rho\Bigr)} ;    
$$
we obtain for quark $u$ and $d$ respectively with $v\,\to\,1$ 
\begin{eqnarray}
& & <p_T^g(u)>\,=\,\frac{9\,\pi\,\Delta\,E_u}{32}\approx\,9.9\,GeV\,;\nonumber\\
& &  <p_T^g(d)>\,=\,\frac{9\,\pi\,\Delta\,E_d}{32}\approx\,4.95\,GeV\,.\label{PT}         
\end{eqnarray}

We have to bear in mind also that in the process of hadronization a gluon give few ordinary hadrons. Their  multiplicity we estimate by the following expression valid in the region of few $GeV$ for charged multiplicity~\cite{N}
\begin{eqnarray}
& &<N_{ch}>\,=\,a\,+\,b\,\ln\sqrt{s}\,;\nonumber\\
& &a\,=\,-\,0.43 \pm 0.09\,;\quad b\,=\,2.75 \pm 0.06\,.\label{nch}         
\end{eqnarray}
Neutral particles have to be also taken into account. 
In view to estimate the total multiplicity we multiply expression~(\ref{nch}) by $\frac{3}{2}$ (isotopic spin of a gluon is zero).  
Then we estimate $\sqrt{s}=\sqrt{2 \Delta E M_p+M_p^2}$ and corresponding mean multiplicity
\begin{eqnarray}
& &u:\;\sqrt{s}\,=\,4.15\,GeV\,;\quad <N>\,=\,5.2\,;\nonumber\\
& &d:\;\sqrt{s}\,=\,3.37\,GeV\,;\quad <N>\,=\,4.3\,.\label{nud}         
\end{eqnarray} 
We take values~(\ref{nud}) with spread $\pm\,2$ and thus obtain estimate for transverse momenta of hadrons $p_T\,=\,p_T^g/N$ 
\begin{eqnarray}
& &u:\;1.3\,GeV\,<\,p_T\,<\,3.0\,GeV\,;\nonumber\\
& &d:\;0.8\,GeV\,<\,p_T\,<\,2.0\,GeV\,.\label{pth} 
\end{eqnarray} 
Estimates~(\ref{pth}) just correspond to the interval of the ridge effect~\cite{CMS}. Of course, by changing light quark masses in their allowable regions we can move boundaries in~(\ref{pth}). However order of magnitude of the effect remains the same.

Next point of our interpretation is that gluons are flying in the narrow cone in the directions of a quark and the angular spread for the multiple gluon radiation is estimated to be
\begin{equation}
\Delta \bar\theta \,\simeq\,\frac{<p_T^g>\,\sqrt{N_g}}{< E_g >\,N_g}\,;\label{dphi1}         
\end{equation}
where $N_g$ is the multiplicity of bremstralung gluons in the event. Obtaining~(\ref{dphi1}) we take into account that average transversal momentum squared for $N_g$ produced gluons 
$$< p_T^2 (N_g)>\, = \,< p_T^g >^2 N_g$$ 
due to statistical nature of the multiple radiation. We take for transversal momentum of a gluon $p_T^g$  estimates~(\ref{PT}) and $< E_g >$ is a mean energy of a gluon. From~(\ref{dphi1}) we see, that for small multiplicity of gluons $\Delta \bar\theta$ increases and this explains why the effect is absent in this case. For estimation of the real experimental situation~\cite{CMS} we replace the denominator in~(\ref{dphi1}) by the energy of partons collision 
 \begin{equation}
\Delta \bar\theta \,\simeq\,\frac{2\,<p_T^g>\,\sqrt{N_g}}{\sqrt{x_1\,x_2\,s}}\,;\label{dphi2}         
\end{equation} 
where  $x_1,\,x_2$ are values of $x$ for quark in the first proton and the anti-quark in the second one.  Number of radiated gluons $N_g$ depends on angle $\psi$ and velocity $v$. Using again formulas from~\cite{LaLi} we have the following estimate
\begin{equation}
N_g\,=\,\frac{\sqrt{x_1 x_2 s}}{2 \Delta E \sqrt{1+\frac{\sin^2\psi}{(1-v^2)}}}\,.\label{NG}
\end{equation} 
For example with $\psi = 0.1$ and $v = 0.999$, average $\Delta E = (\Delta E_u + \Delta E_d)/2 = 8.4\,GeV$, $\sqrt{s}=7\,TeV$~\cite{CMS} and with average of the product $<x_1\,x_2> \approx 0.01$ (see, e.g.~\cite{CTEQ} and references therein) we have 
$N_g \simeq 17$. Bearing in mind, that in our interpretation one bremstralung gluon gives average number of charged hadrons $N_{ch} \approx 3.2$, with $N_g\,\,\approx 17$ we have total number of charged particles produced by a quark $N_{ch}^q = 54$ that gives just multiplicity $\ge 100$ for two radiating quarks. So our mechanism does not contradict to 
real experimental situation~\cite{CMS}. 

Now with $N_g = 17$,   
$\sqrt{s}=7\,TeV$, average $<p_T^g>=7.4\,GeV$  and  $<x_1\,x_2> = 0.01$ we have from~(\ref{dphi2}) 
\begin{equation}
\Delta \bar\theta \,\approx\,0.09\,;\label{dphi3}         
\end{equation}
This angular spread~(\ref{dphi3}) actually gives widening of distributions~(\ref{LLthpsi}, \ref{LLphipsi}) in $\eta$ and $\phi$. The resulting $\Delta \phi$ is to be obtained by simultaneous account of~(\ref{dphi3}) and of $\Phi(\phi)$ width~(\ref{LLphipsi}). Let us also draw attention to widening of the ridge with  $\sqrt{s}$ decreasing. E.g. for $\sqrt{s}\,=\,0.9\,TeV$ in accordance with~(\ref{dphi2}) $\Delta \bar\theta\,=\,0.8$, that means vanishing of the effect.

Thus one can conclude the simple mechanism of gluon bremstralung off
quarks moving in a strong string field describes qualitatively the
CMS ridge effect without adjusting parameters. 
 Of course, a real situation could be much more involved. In
particular, other colour configurations, as was pointed out in various
studies (see, for example,~\cite{ADrem}), may play a significant role. Our
consideration based on simple quasiclassical estimations shows that
constituting string configurations may lead to basic features of the ridge
effect, namely, correlations in particular kinematical region at very high
multiplicities. Obviously, in order to show more accurate properties  of
proposed mechanism one should elaborate in more detail corresponding model
and develop corresponding event generator to perform more realistic simulations.

Let us note, that the accelerated quarks radiate photons as well. The same quasi-classical estimate gives for 
radiated photons two values of mean $p_T$ for two values of (anti-)quark charge (shown in brackets)
 \begin{eqnarray}
& &\biggl(\frac{2\,e}{3}\biggr):\;p_T\,\approx\,2.9\,GeV\,\times\,\frac{2.5}{m_u(MeV)}\,;\nonumber\\
& &\biggl(\frac{e}{3}\biggr):\;p_T\,\approx\,0.72\,GeV\,\times\,\frac{5}{m_d(MeV)}\,.\label{em}         
\end{eqnarray}

It seems to be interesting to check these predictions with CMS data in the region of the ridge. In case of confirmation of the effect, measurement of $p_T$ of the photons may give useful information on current masses of light quarks $m_u,\,m_d$. Let us remind, that for the moment these parameters are known with considerable uncertainty.

The work was supported in part by grant of Russian Ministry of Education
and Science NS-4142.2010.2 and state contract 02.740.11.0244.

%\section{}
%\label{}

%% The Appendices part is started with the command \appendix;
%% appendix sections are then done as normal sections
%% \appendix

%% \section{}
%% \label{}

%% References
%%
%% Following citation commands can be used in the body text:
%% Usage of \cite is as follows:
%%   \cite{key}          ==>>  [#]
%%   \cite[chap. 2]{key} ==>>  [#, chap. 2]
%%   \citet{key}         ==>>  Author [#]

%% References with bibTeX database:

\bibliographystyle{model1a-num-names}
\bibliography{<your-bib-database>}

%% Authors are advised to submit their bibtex database files. They are
%% requested to list a bibtex style file in the manuscript if they do
%% not want to use model1a-num-names.bst.

%% References without bibTeX database:

\end{document}